\documentstyle[epsf,12pt]{article}

\def \vol(#1,#2,#3){{{\bf {#1}} (19{#2}) {#3}}}
\def \NP(#1,#2,#3){Nucl.\ Phys.\          \vol(#1,#2,#3)}
\def \PL(#1,#2,#3){Phys.\ Lett.\          \vol(#1,#2,#3)}
\def \PRL(#1,#2,#3){Phys.\ Rev.\ Lett.\   \vol(#1,#2,#3)}
\def \PRp(#1,#2,#3){Phys.\ Rep.\          \vol(#1,#2,#3)}
\def \PR(#1,#2,#3){Phys.\ Rev.\           \vol(#1,#2,#3)}
\def \PTP(#1,#2,#3){Prog.\ Theor.\ Phys.\ \vol(#1,#2,#3)}
\def \ibid(#1,#2,#3){{\it ibid.}\         \vol(#1,#2,#3)}

\title{Nonmesonic Weak Decay of Light Hypernuclei
       with Coherent $\Sigma$ Mixing}
\author{K.~Sasaki$^{a,b}$, T.~Inoue$^{c}$,  and M.~Oka$^{b}$ \\
 ${^{(a)}}$ 
        Institute of Particle and Nuclear Studies, \\
        High Energy Accelerator Research Organization (KEK), \\ 
        Tsukuba 305-0801, Japan \\
 ${^{(b)}}$ 
        Department of Physics, Tokyo Institute of Technology \\
        Meguro, Tokyo 152-8551 Japan \\
 ${^{(c)}}$ 
        Departamento de F\'{\i}sica Te\'orica and IFIC, \\
        Centro Mixto Universidad de Valencia-CSIC \\
        Institutos de Investigaci\'on de Paterna, \\
        Apdo. correos 22085, 46071, Valencia, Spain}

\begin{document}

\maketitle

\begin{abstract}
    Nonmesonic weak decays of the $A=4$, and 5 hypernuclei are studied.
    The short range parts of the hyperon-nucleon weak interactions are 
    described by the direct quark (DQ) weak transition potential, 
    while the longer range interactions are given by the $\pi$ and $K$ 
    meson exchange processes.
    Virtual $\Sigma$ mixings of the coherent type are found to give 
    significant effects on the decay rates of $^4_{\Lambda}{\rm He}$.
    A large violation of the $\Delta I = 1/2$ rule is predicted
    in the $J=0$ transition amplitudes.
\end{abstract}

\section{Introduction}

Recent progress in experimental studies of hypernuclei has opened a new 
 era of hypernuclear physics that challenges high precision understanding 
 of hypernuclear structures and interactions.
Weak decays of hyperons especially provide us with rich phenomena in 
 which nuclear many-body dynamics
 and nonperturbative QCD effects on hadronic weak interaction
 are important.

Nonmesonic weak decay of hypernucleus is unique as 
 a new type of weak interaction process, which involves only 
 baryons.
In weak decays of heavy hypernuclei, pion emissions via
 $\Lambda\to p\pi^{-}, n\pi^{0}$ are hindered due to the Pauli 
 blocking on the final state nucleon.
Instead, the decay is mainly induced by a two-body transition, 
 $\Lambda N \to NN$, and does not emit pions. 
This nonmesonic weak baryon-baryon interaction
 is analogous to the parity violating part of the nuclear force
 $|\Delta S| =1$ sector.

The mechanism of nonmesonic decay of hypernuclei is a long standing
 problem due to some disagreements between theory and experiment.
One of them is the $\Gamma_{nn}/\Gamma_{pn}$ ratio, i.e.,
 the ratio of $\Gamma_{nn}=\Gamma(\Lambda n \to nn)$
 and $\Gamma_{pn}=\Gamma(\Lambda p \to pn)$ transitions.
Another, to the validity of $\Delta I = 1/2$ rule.
Many studies have been done.
The one-pion-exchange (OPE) is the simplest to describe 
 the $\Lambda N \to NN$ transition.
It has a similar structure as the OPE in nuclear force and in both of 
 them the tensor transition is strong.
As it enhances $\Gamma_{pn}$, the $\Gamma_{nn} / \Gamma_{pn}$ ratio is 
 suppressed.
Typical prediction for OPE is $\Gamma_{nn} / \Gamma_{pn} \sim 0.1$, while 
 most experimental data indicate $\Gamma_{nn} / \Gamma_{pn} \sim 1$ or larger
 ~\cite{Dal:PRL}.

In the nonmesonic decay,
 the $\Lambda$-$N$ mass difference is reflected in a high momentum transfer
 between the baryons, and therefore, short range interaction effects
 must be important.
The lifetime measurements of heavy hypernuclei show saturation 
 at large $A$ and therefore suggest importance of short range interactions
 \cite{Bha:PRL}.

An attempt was made in 
 refs.~\cite{Ose:NPA,Ram01:PRC,Alb:NPA}
 by considering the effect of the polarization 
 of the pion propagator, which seems to enhance $\Gamma_{nn}/\Gamma_{pn}$
 slightly.
The $\rho$ meson contribution was calculated in 
 refs. \cite{McK:PRC,Ban:PTP},
 but it could not reproduce the experimental data.
The two-pion exchange and the effective $\sigma$ meson exchanges
 \cite{Shm:NPA,Ito:NPA,Jid:NPA}
 may play an important role 
 because it increases the central attraction 
 which enhances the $\Gamma_{nn}/\Gamma_{pn}$ ratio through 
 the enhancement of the $J=0$ channels.
The pseudoscalar and vector mesons
 ($\pi$, $K$, $\eta$, $\rho$, $\omega$, and $K^\ast$) 
 are considered in \cite{Dub:AnP,Par:PRC}.
The contribution of each meson comes out large, especially
 for kaon exchange, and significant cancellations between the different
 contributions are found \cite{Dub:NPA}.
The final result is 4-6 times larger than the OPE one
 \cite{Par:hep}.

Apart from the meson exchange model, several authors pointed out 
 the importance of quark degrees of freedom in baryons
 \cite{Che:PRC,Ino01:NPA,Ino02:NPA,Sas:NPA,Mal:PhL}.
Recently we proposed to treat the short range part using the valence
 quark picture of the baryon and the effective four quark weak
 hamiltonian.
We found that the weak quark transition, 
 called the direct quark (DQ) process, gives
 significantly large contribution and shows qualitatively different
 features from the meson exchange mechanism, especially in its
 isospin structure \cite{Ino01:NPA}.
In this process the $\Delta I = 3/2$ contribution is naturally involved, 
 which is found to be important in the $J=0$ decay channels.
We think that DQ is a key process to solve the puzzle of 
 $\Gamma_{nn} / \Gamma_{pn}$ ratio and to reveal the mechanisms of the 
 $\Delta I = 1/2$ rule for the weak $|\Delta S| =1$ decay.

In a previous paper \cite{Sas:NPA}, we proposed a quark-meson hybrid model,
 which includes the DQ transition potential
 supplemented by the long-range part that comes from one-pion (OPE)
 and one-kaon (OKE) exchanges.
It was shown that the model predicts 
 fairly large $\Gamma_{nn} / \Gamma_{pn}$ for the decay of $\Lambda$ in 
nuclear matter. 
In this report, we apply this model to the weak decay of 
 $^4_{\Lambda}{\rm He}$, $^4_{\Lambda}{\rm H}$, and 
 $^5_{\Lambda}{\rm He}$.
We especially concentrate on the effect of virtual $\Sigma$ mixing in 
 $A=4$ system.

In Sect.2, we discuss the basic ingredients of the calculation,
 and the weak transition potential is explained in Sect.3.
The $\Sigma$ mixing contributions are considered in Sect.4.
We show our results in Sect.5, and give the conclusions in Sect.6.

\section{Light hypernuclei}

Light hypernuclei have advantages in order to extract pure information 
 of the elementary process, as the emitted nucleons are
 less distorted than in medium and heavy hypernuclei, as
 $^{12}_{\Lambda}$C, where final state deformations seem to be 
 significant according to the recent data from KEK \cite{Out:PDi}.
Observables of the weak decay of light hypernuclei give us 
 a clue to clear up some puzzles concerning the nonmesonic decay,
 the $\Gamma_{nn}/\Gamma_{pn}$ ratio and the $\Delta I = 1/2$ dominance.
Block and Dalitz~\cite{Dal:PRL} performed
 an analysis based on the lifetime data of light hypernuclei,
 which were updated by some other authors~\cite{Dov:FSS,Sch:NPA}.
They tried to confirm the $\Delta I = 1/2$ dominance
 in medium, but they were unsuccessful so far due to large uncertainties 
 in the data.
Another advantage of the light hypernuclei is that 
 the decay observables may give us evidence of virtual 
 $\Sigma$ excitation
 in $\Lambda$ hypernuclei.
This is the main interest of this paper.

For s-shell hypernuclei,
 the initial $YN$ system can be assumed to be in the relative {\it{s}}-wave 
 state, and we consider the $YN \to NN$ transition with
 the six $^{2S+1}L_J$ combinations listed in Table \ref{TAB:amp}.
Note that $^4_{\Lambda}{\rm He}$ may be mixed with 
 $^4_{\Sigma^{+}}{\rm He}$, in which the $\Sigma^{+}p$ pair induces 
 a new decay channel with the $I^{f}_{z}=+1$ final states, i.e., 
 $\Sigma^{+}p \to pp$.
Thus in Table \ref{TAB:amp}, we have extra amplitudes 
 $a_{pp}$,$b_{pp}$, and $f_{pp}$, which are absent in the previous approaches.
As the two proton final state does not appear
 without the virtual $\Sigma$ state,
 it should give a good signature of the $\Sigma$ mixing in ${^4_\Lambda}$He.

\begin{table}[t]
 \caption{Possible $^{2S+1}L_J$ combination and amplitudes
          for the $YN \to NN$ transitions.}
 \begin{center}
 \begin{tabular}{|rc|ccc|}
 \hline
   $^{2S+1}L_J$ & Final Isospin & \multicolumn{3}{c|}{Amplitudes} \\
     Comb.      & $I^f$         & $I^f_z=0$ & $I^f_z=-1$ & $I^f_z=+1$ 
     \\
 \hline
  $^1S_0   \to  ^1S_0$ & $I^f=1$ & $a_{pn}$  & $a_{nn}$   & $a_{pp}$  
  \\
          $\to  ^3P_0$ & $I^f=1$ & $b_{pn}$  & $b_{nn}$   & $b_{pp}$  
          \\
  $^3S_1   \to  ^3S_1$ & $I^f=0$ & $c_{pn}$  &            &           
  \\
          $\to  ^3D_1$ & $I^f=0$ & $d_{pn}$  &            &           
          \\
          $\to  ^1P_1$ & $I^f=0$ & $e_{pn}$  &            &           
          \\
          $\to  ^3P_1$ & $I^f=1$ & $f_{pn}$  & $f_{nn}$   & $f_{pp}$  
          \\
 \hline
 \end{tabular}
 \end{center}
 \label{TAB:amp}
\end{table}
The main observables are the decay rates.
The total decay rates are the sum of the proton induced $\Gamma_{pn}$
 and the neutron induced $\Gamma_{nn}$ decay rates.
They are given by summing up the squared amplitudes of the relevant
 channels in Table \ref{TAB:amp}.
The decay rates from $J=0$, $\Gamma_{J=0}$, and $J=1$, $\Gamma_{J=1}$,
 channels are often useful, though they are not directly measurable.
The ratio of the parity violating, $\Gamma_{PV}$, and the parity 
 conserving, $\Gamma_{PC}$, decay rates, $PV/PC$, 
 is also an interesting quantity.
Among the six channels given in Table~\ref{TAB:amp},
 $a$, $c$, and $d$ are PC, while $b$, $e$, and $f$ are $PV$.
Although this ratio is not directly observable,
 the asymmetry of the proton emitted from the spin polarized hypernucleus
 is sensitive to $PV/PC$.
The asymmetry parameter is obtained by \cite{Nab:PRC}
 \begin{eqnarray}
  \alpha = \frac{ 2 (\sqrt{3}[a e] - [b c] + \sqrt{2}[b d] 
    + \sqrt{6} [c f] + \sqrt{3}[d f] ) }
      { |a|^2 + |b|^2 + 3\left( |c|^2 + |d|^2 + |e|^2 +|f|^2 \right) }
  \label{EQ:asy}
 \end{eqnarray}
where we define $[ae] \equiv {\rm{Re}} (a^\ast_{pn} e_{pn})$, etc.
Note that there appear interference terms between 
 the $J=0$ and $J=1$ amplitudes, 
 such as $[ae]$ and $[bc]$, in eq.(\ref{EQ:asy}).
The previous calculations often neglected these interference terms,
 but they are important because their magnitudes 
 are similar to the other terms.

The wave functions of the {\it{s}}-shell hypernuclei are rather simple.  
We assume that the nucleons reside in the lowest energy state of the 
 harmonic oscillator shell model, i.e., given by $(0s)^{n}$ 
 configuration. 
The size parameter is chosen so as to reproduce the size of the nucleus 
 without $\Lambda$.  
Recent theoretical and experimental studies suggest 
 that the size of the nucleus shrinks due to the intruded $\Lambda$
 \cite{hiy:npa}, 
 but here we take a conservative approach. 
Calling the $(0s)^{n}$ nucleons as the core, the $\Lambda$-core 
 relative motion is described by the solution of the Schr\"odinger equation 
 with a $\Lambda$-core potential obtained by the convolution of the 
 realistic $\Lambda-N$ interaction.

It was shown that the short-range repulsion between $\Lambda$ and $N$ 
 results in a repulsion at the center of the core and thus the $\Lambda$ 
 is pushed out from the core region.  
Such a wave function was shown to explain the mesonic decay rates 
 of the light hypernuclei, which are sensitive to the overlap of the $\Lambda$ 
 and $N$ wave functions \cite{Mot:NPA}.
When we consider the virtual $\Sigma$ mixing,  
 we assume that the $\Sigma$ single particle wave function is given by
 a Gaussian, whose $b_\Sigma$ parameter is adjusted according to 
 the $\Sigma$ mass,
\begin{eqnarray}
 b_\Sigma = \sqrt{\frac{M_\Sigma + M_N}{M_\Sigma}} b_N
\end{eqnarray}
Thus $\Sigma$ resides more in the central region than $\Lambda$ and
 its effect on the weak decay is enhanced.
This enhancement is about 10\% in magnitude, as can be shown by comparing
 with the calculation assuming that $\Sigma$ wave function is
 identical to $\Lambda$.

In computing the two-body decay matrix elements, it is necessary 
 to take into account the short-range correlation.  
Thus we take the wave function of the Y-N two body systems in the form,
\begin{eqnarray}
  \phi_Y({\vec{r}}_Y) \phi_N({\vec{r}}_N)
      \left[
        \left(
        1 - e^{-r^2 / a^2}
        \right)^n
      - br^2 e^{-r^2 / c^2}
      \right]
\end{eqnarray}
 with $r=|\vec r_Y - \vec r_N |$ and determine the parameters
 for the SRC so that it reproduces the realistic $\Lambda N$ 
 correlation \cite{Par01:PRC},
 which gives $a= 0.5$, $ b=0.25$, $c=1.28$ and $n=2$.

The wave function of the final two nucleons emitted in the two-body 
 weak process is assumed to be the plane wave with SRC:
\begin{eqnarray}
 e^{i \vec K \cdot \vec R'} e^{i \vec k \cdot \vec r'}
        \left[
        1 - j_0(q_c r')
        \right]
\end{eqnarray}
 where $\vec r'=\vec r_{N_2} - \vec r_{N_1}$,
 $\vec R'= (\vec r_{N_2} + \vec r_{N_1})/2$
 and $q_c$ = 3.93 [fm$^{-1}$] .
This approximation may be justified for light nuclei as the 
 momenta of the emitted nucleons are relatively high ($\sim 400$ MeV/c).

\section{Transition potential}

We employ a hybrid model to describe the weak
 $YN \leftrightarrow NN$ transition potential
 \cite{Ino01:NPA,Ino02:NPA,Sas:NPA}.
At long and medium distances, the transition is induced mainly by
the one pion exchange (OPE) and one kaon exchange (OKE)
mechanisms.
For example the $\Lambda p \to np$ transition potential
 induced by $\pi^0$ exchange is given by
\begin{eqnarray}
 \lefteqn{V_{\Lambda p \to np}(\vec q)} \nonumber \\
  && =
  G_F m_{\pi}^2
    [\bar u_n (A^\Lambda_{\pi}+B^\Lambda_{\pi} \gamma_5) u_{\Lambda}]
    {\frac{I_1 \times I_2}{{\vec q}^2 + {\tilde m}_{\pi}^2}}
    \left( {  \frac{ \Lambda_{\pi}^2 -{\tilde m}_{\pi}^2 }
              { \Lambda_{\pi}^2 + {\vec q}^2   }
     } \right)^2
  g_{\pi NN} [\bar u_p \gamma_5 u_p]
  \label{eq:OPE}
\end{eqnarray}
 where
 $I_1$ and $I_2$ are the isospin factors and, in this case, given as
\begin{eqnarray}
 I_1 = \langle n | \tau^3_1 | \Lambda \rangle = -1, ~ ~ ~
 I_2 = \langle p | \tau^3_2 | p \rangle = 1.
\end{eqnarray}
Here $\Lambda$ is regarded as a $|I ,I_3 \rangle = |1/2,-1/2 \rangle$
 state, called spurion.
This form guarantees that this transition is purely given by
 $\Delta I = 1/2$ amplitude.
The coupling constants $g_{\pi NN}$, $G_F$, $A_{\pi}$ and $B_{\pi}$
 are determined phenomenologically so that the free $\Lambda$ decay rate
 is reproduced.
A double pole form factor with the cutoff
 parameter $\Lambda_{\pi}$ = 800 MeV is employed.
As the energy transfer is significantly large 
 we introduce the effective pion mass
\begin{eqnarray}
\tilde{m}_\pi = \sqrt{m_\pi^2 - (m_\Lambda - m_N)^2/4}
 \simeq 110 {\rm{MeV}}.
\end{eqnarray}

The Dirac spinors and the $\gamma$ matrices in eq.(\ref{eq:OPE}) are
 reduced into the nonrelativistic form in the standard way.
After the spin and angular momentum projections, we obtain
 a local potential for each transition channel in Table
 \ref{TAB:amp}.
The cut off $\Lambda_{\pi}$ = 800 MeV for OPE is rather soft compared
 to a harder cut off ($\Lambda_{\pi}$ = 1300 MeV)
 employed by the one-boson exchange (OBE) potential
 model of nuclear force \cite{Mac:PhR}.
On the other hand, the soft form factor is preferred in the study
 of $NN \to NN \pi$ process \cite{Lee:PRC}.
A reason for this discrepancy is that the OBE potential model requires
 reasonably strong repulsion and spin-dependent forces induced
 by vector meson exchange.
As the vector mesons are heavy, they need a hard form factor.
In our approach, however, the short distance part is described in terms
 of quark substructures of baryons and only the pion and kaon are employed
 for the meson exchange potential.
We therefore believe that the soft form factor is more consistent for
 our calculation.

The kaon exchange potential (OKE) can be constructed similarly.
Both the strong and weak coupling constants are evaluated under the
 assumption of the flavor SU(3) symmetry.
The cut off for the kaon vertex is taken as $\Lambda_{K}$ = 1200 MeV,
 according to \cite{Par:PRC}.
All the couplings used in our calculation are listed in Table \ref{TAB:cc}.
Note that, for OKE, it involves the strangeness transfer and thus
 the strong and weak vertices are exchanged.

\begin{table}[t]
 \caption{The strong and weak meson baryon coupling constants.
          The weak coupling constants are given
          in units of $G_F m_\pi^2 =2.21 \times 10^{-7}$.}
 \begin{center}
 \begin{tabular}{|c||l|l|l|c|}
 \hline
 Meson  & \multicolumn{1}{c|}{Strong c.c.}
        & \multicolumn{2}{c|}{Weak c.c.} & Cut off $\Lambda$
 [MeV] \\
 \cline{3-4}
  & & \multicolumn{1}{c|}{PC} & \multicolumn{1}{c|}{PV} & \\
 \hline
 $\pi$  & $g_{NN\pi}=13.3$  & $B^\Lambda_\pi=-7.15$  & $A^\Lambda_\pi=1.05$
&
800  \\
        & $g_{\Lambda \Sigma \pi}=12.0$ &
            $B^{\Sigma^+}_{\pi^+}=-18.3$ &
            $A^{\Sigma^+}_{\pi^+}= 0.04$ &
\\
        & & $B^{\Sigma^+}_{\pi^0}=-12.2$ &
            $A^{\Sigma^+}_{\pi^0}=-1.39$ &
\\
        & & $B^{\Sigma^0}_{\pi^0}=-8.78$ &
            $A^{\Sigma^0}_{\pi^0}=0.95$ &
\\
        & & $B^{\Sigma^0}_{\pi^-}=12.2$ &
            $A^{\Sigma^0}_{\pi^-}=1.39$ &
\\
        & & $B^{\Sigma^-}_{\pi^-}=0.74$ &
            $A^{\Sigma^-}_{\pi^-}=1.87$ &
\\
 \hline
 $K$    & $g_{\Lambda NK}=-14.1$ & $C_K^{PC}=-18.9$ & $C_K^{PV}=0.76$ &
1200 \\
        & $g_{\Sigma NK}=4.28$  & $D_K^{PC}=6.63$   & $D_K^{PV}=2.09$ &
\\
 \hline
 \end{tabular}
 \end{center}
 \label{TAB:cc}
\end{table}

For shorter distances, we employ the direct quark (DQ) transition 
potential based on the constituent quark picture of baryons.  In this 
mechanism, a strangeness changing weak interaction between two 
constituent quarks induces the transition of two baryons.  
Here we sketch the derivation, while the details are given in
 \cite{Ino01:NPA}.
 
The DQ transition potential is derived by evaluating the 
matrix elements,
 \begin{eqnarray}
    V_{DQ}(k,k')_{{L_i,S_i,J}\atop{L_f,S_f,J}} \equiv \langle 
    NN(k',L_f,S_f,J) | H_{eff}^{\Delta S=1} | YN(k,L_i,S_i,J)\rangle.
 \end{eqnarray}
Here the two-baryon states, $| BB(k,L,S,J) \rangle$, are
expressed by six-quark wave functions, constructed in the quark 
cluster formalism.  As we use nonrelativistic quark model,
we employ the transition potential given by nonrelativistic reduction 
of the low-energy effective weak Hamiltonian for $|\Delta S|=1$, 
consisting of 4-quark weak vertices:
\begin{equation}
   H_{eff}^{\Delta S=1} = 
  -\frac{G_f}{\sqrt 2}\sum_{r=1,r\ne 4}^6K_r O_r
\label{eqn:heff}
\end{equation}
where $O_{r}$'s are the 4-quark operators whose explicit forms
are given below.

This Hamiltonian is derived by the renormalization-group-improved 
perturbation theory of QCD from the standard electro-weak vertex,
\begin{eqnarray}
 \left\{
  \begin{array}{l}
   s \to u + W^- \\
   u + W^- \to d
  \end{array}
 \right.
\end{eqnarray}
 or
\begin{eqnarray}
 s + u \to u + d.
\label{EQ:weakqt}
\end{eqnarray}
Note that eq.(\ref{EQ:weakqt}) contains both $\Delta I = 1/2$ and $3/2$
contributions with similar strengths.
It is known that the QCD correction enhances the $\Delta I = 1/2$ 
part, while it suppresses the $\Delta I = 3/2$ part at the same time
 \cite{Gai:PRL,Alt:PhL}.
The mechanism of $\Delta I = 1/2$ enhancement can be intuitively
 understood by considering one gluon exchange between quarks
 in the initial or final two quark states.
The spin dependent part of the one gluon exchange interaction
 lowers the energy of color antisymmetric $J=0$ pair, which
 restricts the final state to $I=0$.
Therefore the $\Delta I = 1/2$ transition from $I_i = 1/2$ to
 $I_f = 0$ is enhanced.
Further enhancement comes from the so-called penguin diagrams,
 which induce the $O_5$ and $O_6$ operators given below
 \cite{Vai:JETP}.

We employ the values of the coefficients $K_{r}$ 
evaluated by 
solving the renormalization-group equations\cite{Gil:PRD,Pas:NPB},
\begin{eqnarray}
  \begin{array}{c|c|c|c|c}
  \quad  K_1   \quad  &   \quad  K_2   \quad
  &
  \quad  K_3   \quad  &   \quad  K_5   \quad
  &
  \quad  K_6   \quad \\
 \hline
   -0.284 &  0.009  & 0.026 & 0.004 & -0.021 \\
  \end{array}
  \nonumber
\end{eqnarray}
each of which correspond to the four-quark operators,
\begin{eqnarray}
  O_1 &=& (\bar d_{\alpha}s_{\alpha})_{V-A}
          (\bar u_{\beta}u_{\beta})_{V-A}
         -(\bar u_{\alpha}s_{\alpha})_{V-A}
           (\bar d_{\beta}u_{\beta})_{V-A}
  \nonumber\\
  O_2 &=& (\bar d_{\alpha}s_{\alpha})_{V-A}
           (\bar u_{\beta}u_{\beta})_{V-A}
         +(\bar u_{\alpha}s_{\alpha})_{V-A}
           (\bar d_{\beta}u_{\beta})_{V-A}
  \nonumber\\ 
     & &+2(\bar d_{\alpha}s_{\alpha})_{V-A}
         (\bar d_{\beta}d_{\beta})_{V-A}
        +2(\bar d_{\alpha}s_{\alpha})_{V-A}
         (\bar s_{\beta}s_{\beta})_{V-A}
  \nonumber\\  
  O_3 &=& 2(\bar d_{\alpha}s_{\alpha})_{V-A}
           (\bar u_{\beta}u_{\beta})_{V-A}
         +2(\bar u_{\alpha}s_{\alpha})_{V-A}
           (\bar d_{\beta}u_{\beta})_{V-A}
  \\
     & &-(\bar d_{\alpha}s_{\alpha})_{V-A}
         (\bar d_{\beta}d_{\beta})_{V-A}
        -(\bar d_{\alpha}s_{\alpha})_{V-A}
         (\bar s_{\beta}s_{\beta})_{V-A}
  \nonumber\\
  O_5 &=& (\bar d_{\alpha}s_{\alpha})_{V-A}
       (\bar u_{\beta}u_{\beta}+\bar d_{\beta}d_{\beta}
      + \bar s_{\beta}s_{\beta})_{V+A}
    \nonumber\\
  O_6 &=& (\bar d_{\alpha}s_{\beta})_{V-A}
       (\bar u_{\beta}u_{\alpha}+\bar d_{\beta}d_{\alpha}
      + \bar s_{\beta}s_{\alpha})_{V+A}
  \nonumber
\end{eqnarray}
Among the above four-quark operators, only $O_{3}$ induces $\Delta
I=3/2$ 
transitions. The large value of $K_{1}$, and the appearance of $K_{5}$ 
and $K_{6}$ (from the penguin diagrams) show the enhancement of
$\Delta I=1/2$ transition.

It is, however, realized that the above perturbative enhancement is
not 
enough to explain the observed $\Delta I=1/2$ dominance in the decays 
of the kaon and the hyperons.  Further enhancement, most probably 
due to nonperturbative QCD corrections, is required.
Several possibilities have been suggested, which include effects of 
isospin dependent final state interactions for the 
$K \to \pi \pi$ decays\ \cite{Mor:PRL,Ino03:PTP},
and for the hyperon decays, the suppression of the $\Delta I=3/2$ by 
the color antisymmetrization of the valence
 quarks \cite{Miu:PTP,Pat:PRD}.  
Neither of them seem to be effective in the two-baryon transitions 
in the present analysis.  Thus it is possible that the nonmesonic 
weak decays show significant deviation from the $\Delta I=1/2$ dominance.

\section{$\Lambda$-$\Sigma$ mixing contribution}

Virtual $\Sigma$ can be mixed in $\Lambda$ hypernuclei
 via the strong $\Lambda N \to \Sigma N$ transition.
Effects of $\Sigma$ mixing has been considered by many authors
 in the context of both the hypernuclear structure and its transitions.

Recently, it is advocated that the $\Sigma$-mixing is crucial in 
 solving the overbinding problem of the {\it{s}}-shell hypernuclei
 \cite{Aka:PRL}.  
Namely, the coherent $\Sigma$ mixing, which is important in $A=4$ 
 hypernuclei, gives enough attraction for the $A=4$ binding energy 
 even if we take weaker central attraction that is preferable for the 
 smaller binding of $^{5}_{\Lambda}$He.
A sophisticated four-body calculation of $A=4$ hypernuclear structure 
 also indicates significant mixing of virtual $\Sigma$ of 1-2\% level 
 and thus supports the above idea\cite{hiy:npa}.

If the mixing probability of the virtual $\Sigma$ is 1\%, the mixing 
 amplitude $\beta$ is $| \beta | \sim 0.1$.
Although its effects on the binding energy are proportional to 
 $|\beta|^{2}$ perturbatively, those on the transition amplitude are of 
 the order of $| \beta |$.  
The latter is also sensitive to the phase of the mixing 
 and therefore to the mixing mechanism.
Thus we consider the coherent $\Sigma$ mixing in  nonmesonic decays
 of the $A=4$ hypernuclei.

\begin{figure}[t]
 \centerline{ \epsfxsize=4cm \epsfbox{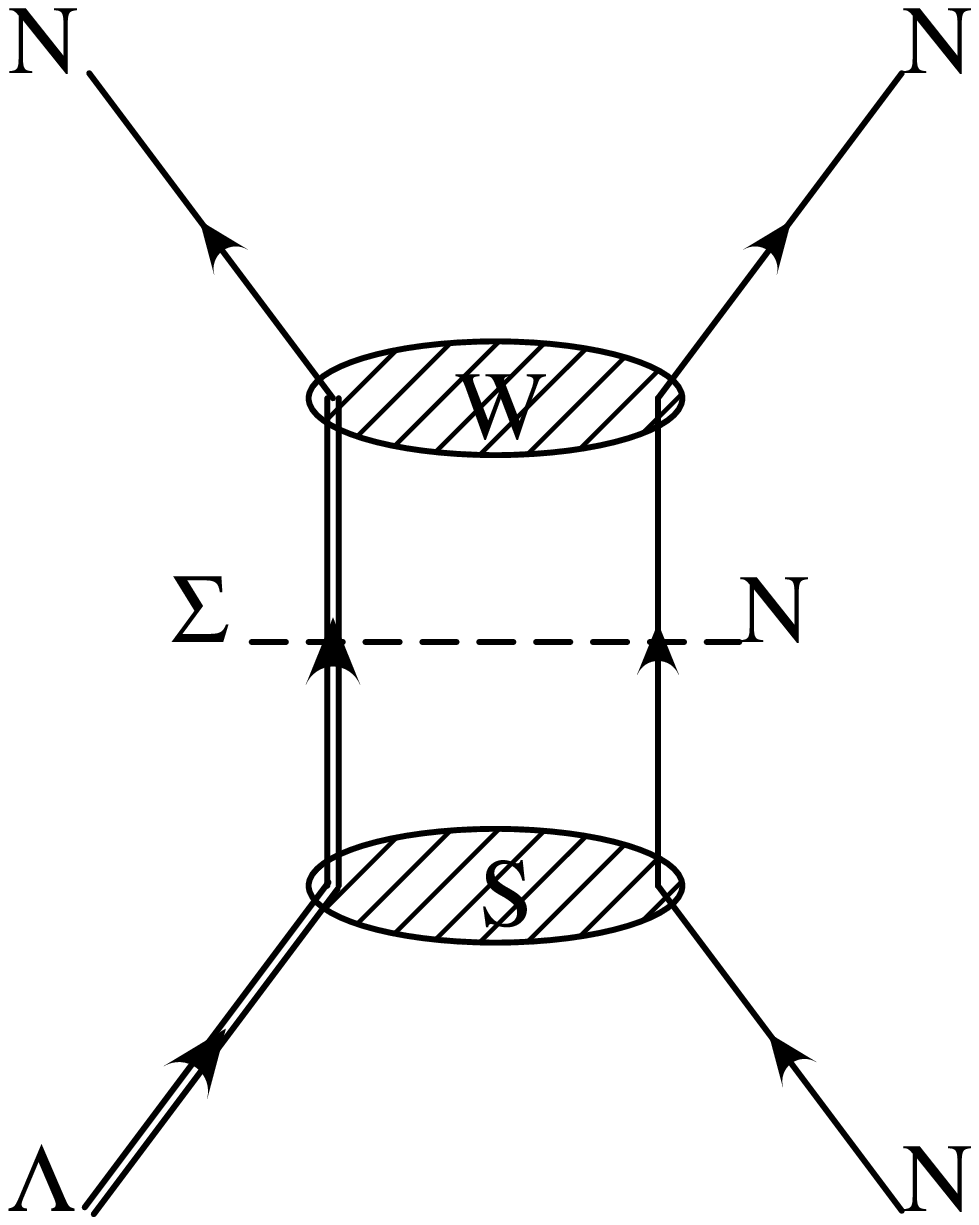} ~{\large{(a)}}
              \hspace*{2cm}
              \epsfxsize=5.5cm \epsfbox{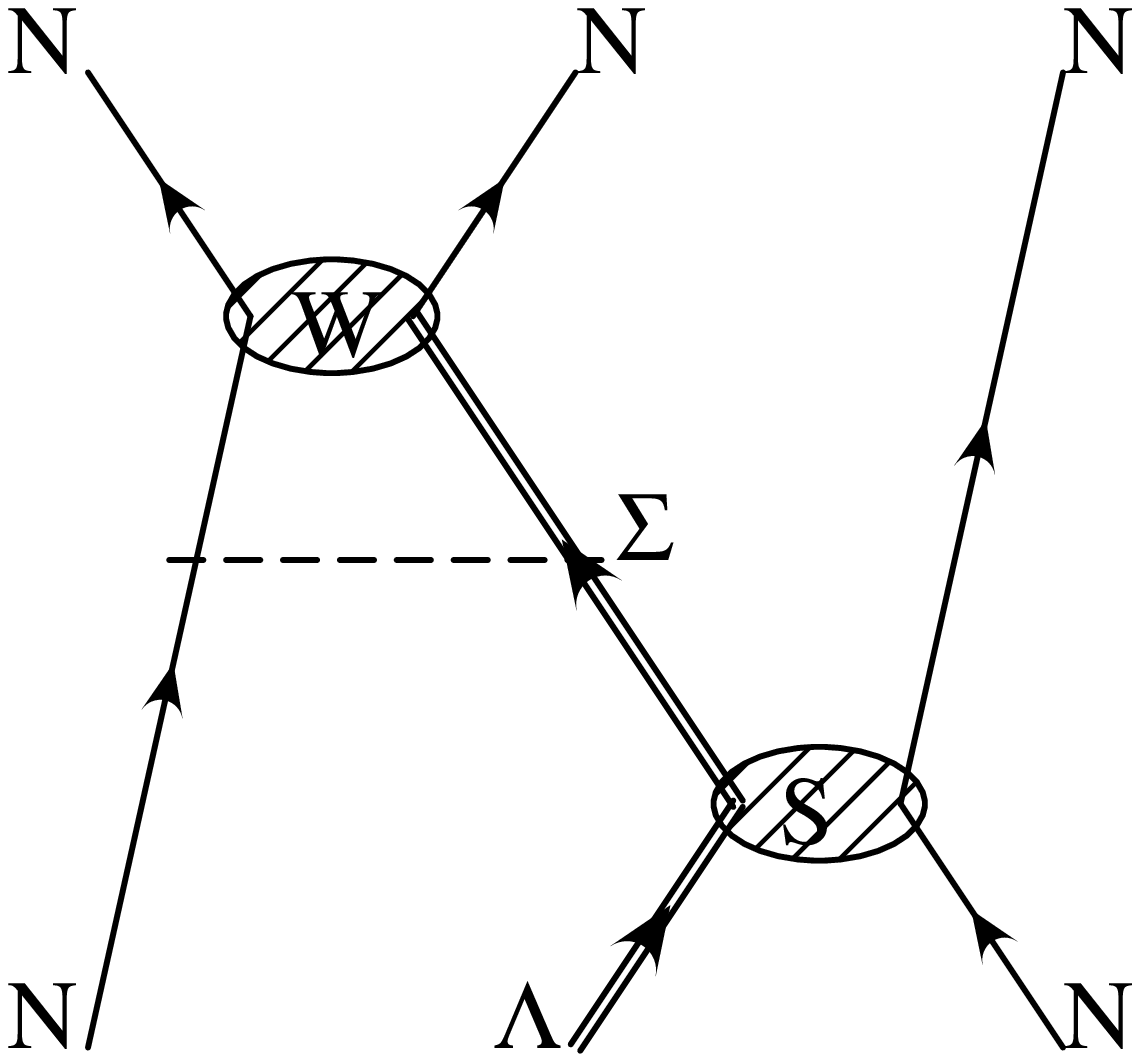} ~{\large{(b)}} }
\caption{Diagrammatic representation of two and three body type
         $\Sigma$ mixing for the $\Lambda N \to NN$ transition.}
  \label{sig}
 \end{figure}

The diagrams shown in fig.\ref{sig} are two types of nonmesonic weak 
 decays of the virtual $\Sigma$ in nuclei.
A previous study \cite{ban:ijm} considered two-pion exchange process
 between $\Lambda$ and N, one of which induces weak transition
 (fig.\ref{sig}(a)).
The intermediate $\Sigma$-N state is restricted to $I=1/2$.
It is, however, possible that the virtual $\Sigma$ decays
 with the assistance of a second nucleon, (fig.\ref{sig}(b)).
This ``three body'' type process is taken into account by considering 
 the coherent $\Sigma$ mixing.  They are important for two reasons.
\begin{description}
\item{(1)}
    It involves the weak interaction of the
    $\Sigma$-N ($I=3/2$) states,
    which does not contribute in fig.\ref{sig}(a).
\item{(2)} 
    The coherent mixing of $\Sigma$ hypernuclear states
    is prohibited in a hypernucleus with $I=0$,
    such as ${^5_\Lambda}$He, due to the isospin conservation.
    In contrast, for $I \neq 0 $ hypernuclei, the coherent 
    $\Sigma$ mixing allows the virtual $\Sigma$ to interact 
    with all the nucleons equally and therefore the 3-body weak process
    fig.\ref{sig}(b) becomes important.
\end{description}

Let us write down the $\Sigma$ mixing effect explicitly 
 for ${^4_\Lambda}$He.
We suppose that the $\Sigma$+3N state with the same quantum numbers
 mixes to the $|\Lambda+{^3}{\rm{He}} \rangle$ state,
\begin{eqnarray}
| {^4_\Lambda}{\rm{He}} \rangle^0_{1/2}
          = \sqrt{ 1 - \beta^2 } | \Lambda + {^3}{\rm{He}}
\rangle^0_{1/2}
            +
            \beta | \Sigma + {\rm{3N}} \rangle^0_{1/2}
\end{eqnarray}
 where the superscripts and subscripts are the total angular momentum
 $(J)$ and isospin $(I)$ respectively and
\begin{eqnarray}
 \lefteqn{ | \Sigma + 3N \rangle^0_{1/2} } 
 \nonumber \\
 ~\ ~\
  &=&
   \sqrt{ \frac{2}{9} } | (\Sigma^+ p)^0 (nn)^0 \rangle^0_{1/2}
   +
   \sqrt{ \frac{1}{9} } | (\Sigma^0 n)^0 (pp)^0 \rangle^0_{1/2}
 \nonumber \\
  & & ~\ ~\ ~\ ~\ ~\ ~\ ~\ 
   -
   \sqrt{ \frac{1}{6} } | 
   [ \sqrt{ \frac{2}{3} } (\Sigma^+ n)^0 
   + \sqrt{ \frac{1}{3} } (\Sigma^0 p)^0 ] (pn)^0  \rangle^0_{1/2}
 \nonumber \\
  & & ~\ ~\ ~\ ~\ ~\ ~\ ~\ ~\ ~\ ~\ ~\ ~\
   +
   \sqrt{ \frac{1}{2} } | 
   [ \sqrt{ \frac{2}{3} } (\Sigma^+ n)^1 
   - \sqrt{ \frac{1}{3} } (\Sigma^0 p)^1 ] (pn)^1  \rangle^0_{1/2}.
\hspace*{2.5cm} 
\end{eqnarray}
One sees that $\Sigma + 3N$ state contains the $\Sigma^+ p$ $(J=0)$
 component with 22\% probability for each pair of baryons. 
The $\Sigma$ mixing strength is denoted by $\beta$.
According to the recent calculation by Hiyama {\it{et al.}}\cite{hiy:npa},
 $|\beta|^2$ is about 1\% in the ground state of ${^4_\Lambda}$He.
But we need not only the magnitude but also the relative
 phase between the $\Lambda + 3{\rm N}$ and $\Sigma + 3{\rm N}$ states.
We therefore attempt to estimate $\beta$ roughly in the first order 
 perturbation as follows
\begin{eqnarray}
\beta = -
 \frac{
 \langle \Sigma {\rm +}{^3}{\rm{He}} | 
 V_{\Sigma N \to \Lambda N} 
 | \Lambda{\rm +}{^3}{\rm{He}} \rangle
      }  {M_\Sigma - M_\Lambda}
\label{eq:pert}
\end{eqnarray}
For the transition potential, we employ the D2 potential of the paper
 \cite{Aka:PRL}. 
Evaluating eq.(\ref{eq:pert}) by using the Gaussian wave function,
 we obtain
\begin{eqnarray}
 \beta = -0.05 
\end{eqnarray}
We should note that the magnitude of $\beta$ is rather sensitive 
 to the Gaussian $b$ parameter, which is chosen here as $b=2.24$ [fm].
Although this estimate of $\beta$ may be too crude to be quantitative, 
 we can at least determine the sign of the mixing.  
In the next section, 
 we assume the coherent $\Sigma$ mixing of 1\% for ${^4_\Lambda}$He
 and evaluate its effects on the nonmesonic decay rates.
We consider the similar $\Sigma$ mixing 
 to the isospin partner ${^4_\Lambda}$H.

The transition potential for the $\Sigma N \to NN$ is derived 
 similarly to the $\Lambda N \to NN$ transition.
For example the OPE induced $\Sigma^0 p \to np$ transition potential is
\begin{eqnarray}
 \lefteqn{V_{\Sigma^0 p \to np}(\vec q)} \nonumber \\
  &&=
  G_F m_{\pi}^2
    [\bar u_n
     (A^{\Sigma^0}_{\pi^0}+B^{\Sigma^0}_{\pi^0} \gamma_5) 
    u_{\Sigma^0}]
    {\frac{I_2}{{\vec q}^2 + {\tilde m}_{\pi}^2}}
    \left( {  \frac{ \Lambda_{\pi}^2 -{\tilde m}_{\pi}^2 }
              { \Lambda_{\pi}^2 + {\vec q}^2   }
     } \right)^2
  g_{\pi NN} [\bar u_p \gamma_5 u_p]
  \label{eq:OPEs}
\end{eqnarray}
where $I_2 = 1$.
The potential for the OKE induced $\Sigma N \to NN$ transition is also
 written similarly.

\section{Results}

\subsection{A=5 system}

Table \ref{TAB:He5} shows the results for ${^5_\Lambda}$He.
The $\pi$ exchange process has a large proton-induced decay rate $\Gamma_{pn}$
because of the large tensor transition amplitude ($d_{pn}$-channel).
The $K$ exchange reduces the $\pi$ contribution of $\Gamma_{pn}$ by about 
30 \%,
while it enhances the $J=1$ part of $\Gamma_{nn}$
($f_{nn}$-channel) and the $\Gamma_{nn}/\Gamma_{pn}$ ratio.

The DQ process gives significantly large contribution
and large $\Gamma_{nn}/\Gamma_{pn}$ ratio,
which is the major difference from the meson exchange.
It contributes constructively to the meson exchange amplitudes.
As a result, we have $\Gamma_{nn}/\Gamma_{pn}$ ratio largely enhanced to 
$0.720$.  Considering the large error bars in the experimental data, 
the agreement with experiment is fairly good.  

Exchanges of the heavy mesons 
 i.e. $\eta$, $\rho$, $\omega$, and $K^\ast$,
 are the other possibility for the short range weak interaction.
It was pointed out that the nonmesonic decay rates are significantly
 enhanced by the heavy mesons,
 while the change of $\Gamma_{nn}/\Gamma_{pn}$ ratio is not large 
\cite{Par:hep}.
In contrast, the DQ process induces a large $\Gamma_{nn}$ and it gives
 $\Gamma_{nn}/\Gamma_{pn} \simeq 1.216$ by itself.
From the $\pi + K$ exchange contribution 
 the $\Gamma_{nn}/\Gamma_{pn}$ ratio is enhanced by more than 50\% 
 due to the DQ transition.
Ref.\cite{Par:hep} examined the realistic final state interactions and
 found a stronger effect that reduces the total nonmesonic decay rates
 by more than 50\% and the $\Gamma_{nn}/\Gamma_{pn}$ ratio by more than
 20\%.
As the recent measurement of the nucleon spectrum \cite{Oka:pdi}
 suggests significantly large final state interaction,
 further quantitative studies are important and urgent.

The last column of Table \ref{TAB:He5} shows the asymmetry parameter
 of the emitted proton against the polarization of $\Lambda$. 
This quantity is sensitive to the ratio of parity violating amplitude
 and parity conserving amplitude. 
Our result shows that the asymmetry parameter is negative and large
 as it is enhanced by the DQ contribution compared to the value given by
 OPE.
This again is a reflection of strong $f_{pn}$ amplitude.  
The only available experimental data taken at KEK indicate positive 
 value in contrast to the theoretical predictions \cite{aji:prl}.  
This requires 
 further study as the sign of the asymmetry seems robust in theoretical 
 calculations as far as the meson exchange and direct quark processes 
 are concerned.

\begin{table}[htb]
\caption{Nonmesonic decay rates of ${^5_\Lambda}$He
         in units of $\Gamma_\Lambda$ }
\begin{center}
\begin{tabular}{p{1.9cm}p{1,8cm}p{1.8cm}p{1.8cm}p{1.8cm}p{1.8cm}p{1.8cm}}
\hline
${^5_\Lambda}$He
& total & $\Gamma_{pn}$ & $\Gamma_{nn}$ &
$\Gamma_{nn} / \Gamma_{pn}$ & PV/PC & $\alpha$ \\
\hline
$\pi$      & 0.372 & 0.328 & 0.044 & 0.133 & 0.481 & -0.441 \\
$\pi$+$K$    & 0.304 & 0.207 & 0.097 & 0.466 & 1.336 & -0.362 \\
DQ         & 0.066 & 0.030 & 0.036 & 1.216 & 4.403 & -0.398 \\
$\pi$+$K$+DQ & 0.523 & 0.304 & 0.219 & 0.720 & 4.845 & -0.678 \\
\hline
EXP \cite{jjs:prc}
           & 0.41$\pm$0.14 & 0.21$\pm$0.07 &
                      0.20$\pm$0.11 & 0.93$\pm$0.55
           & ----- & ----- \\
EXP \cite{hno:psg}
           & 0.50$\pm$0.07 & 0.17$\pm$0.04 &
                      0.33$\pm$0.04 & 1.97$\pm$0.67
           & ----- & ----- \\
EXP \cite{aji:prl}
           & ----- & ----- & ----- & -----
           & ----- & 0.24$\pm$0.22 \\
\hline
\end{tabular}
\end{center}
\label{TAB:He5}
\end{table}

\subsection{A=4 system}

An advantage of the $A=4$ and 5 hypernuclei regarding the nonmesonic weak 
 decay is their selectivity of the two-body channels.
To the leading order, the ground state of ${^4_\Lambda}$He contains 
 two protons forming spin 0 and a neutron and a $\Lambda$ forming spin 0.
Thus there is no $\Lambda$-$n$ pair with spin 1, which forbids
 the $f_{nn}$-channel in Table \ref{TAB:amp}.
As a result, the neutron induced decay rate, $\Gamma_{nn}$, is strongly 
 suppressed, which is consistent with what current experimental data
 indicate ( Table \ref{TAB:He4}).

It is interesting to see how large the effect of the $\Sigma$ mixing is.
As we discussed in the previous section,
 we consider virtual $\Sigma$-hypernucleus component,
 whose probability is given by $|\beta|^2$.
In Table \ref{TAB:He4}, we show the decay rates of A=4 hypernuclei
 for two values of $\beta$, such that $\beta^{2}\sim 1\%$.
One sees that the mixing changes the total decay rate by about 20\% 
 for $\beta^{2}\sim 1\%$.
The sign of $\beta$ determined above is negative.
One sees that the results for $\beta = -0.1$ give better agreement
 with the current experimental data.
It has been also found that the negative $\beta$ is consistent
 with the study of the magnetic moments
 due to the $\Sigma$ mixing in hypernuclei \cite{Oka:Pro}.
The negative $\beta$ is found to reduce the proton induced decay rate, 
 $\Gamma_{pn}$, while a positive $\beta$ will enhance the rates.
We find that the mixing does not change the 
 qualitative behaviors of the $\Gamma_{nn}/\Gamma_{pn}$ ratio 
 and the proton asymmetry.

A new interesting decay channel is available when we consider the 
 $\Sigma^{+}$ mixing in ${^4_\Lambda}$He.  The system consists of a 
 virtual $\Sigma^{+}$, $p$ and two $n$.  When $\Sigma^{+}$ meets the 
 proton, it decays into two $p$, i.e., 
 $\Sigma^{+} p \to pp$ decay. 
The calculated decay rate for $\Sigma^{+} p \to pp$ is
\begin{eqnarray}
 \Gamma_{pp} = 0.0003 \Gamma_\Lambda
\end{eqnarray}
at $\beta =0.1$. 
Although the branching ratio is tiny, it gives a clean signal as 
 a back to back $p$-$p$ in the final state. 
This will be a direct evidence 
 for virtual $\Sigma$ mixing in $\Lambda$ hypernuclei.
\begin{table}[htb]
\caption{Nonmesonic decay rates of ${^4_\Lambda}$He
         in units of $\Gamma_\Lambda$ }
\begin{center}
\begin{tabular}{p{1.9cm}p{1.8cm}p{1.8cm}p{1.8cm}p{1.8cm}p{1.8cm}p{1.8cm}}
\hline
${^4_\Lambda}$He
& total & $\Gamma_{pn}$ & $\Gamma_{nn}$ &
$\Gamma_{nn} / \Gamma_{pn}$ & PV/PC & $\alpha$ \\
\hline
$\pi$      & 0.272 & 0.250 & 0.022 & 0.089 & 0.353 & -0.417 \\
$\pi$+$K$    & 0.155 & 0.145 & 0.009 & 0.064 & 0.146 & -0.357 \\
DQ         & 0.032 & 0.021 & 0.011 & 0.516 & 2.093 & -0.373 \\
$\pi$+$K$+DQ & 0.218 & 0.214 & 0.004 & 0.019 & 2.321 & -0.679 \\
\hline
~\ $\beta =  +0.1$
           & 0.276 & 0.270 & 0.006 & 0.021 & ----- & -0.678 \\
~\ $\beta =  -0.1$
           & 0.168 & 0.165 & 0.003 & 0.017 & ----- & -0.645 \\
\hline
EXP \cite{Zep:NPA}
           & 0.20$\pm$0.03 & 0.16$\pm$0.02
                    & 0.04$\pm$0.02 & 0.25${^{+0.05}_{-0.13}}$
           & ----- & ----- \\
EXP \cite{Out:NPA}
           & 0.17$\pm$0.05 & 0.16$\pm$0.02
                    & 0.01${^{+0.04}_{-0.01}}$ 
                    & 0.06${^{+0.28}_{-0.06}}$
           & ----- & ----- \\
\hline
\end{tabular}
\end{center}
\label{TAB:He4}
\end{table}

The selectivity is reversed in ${^4_\Lambda}$H, that is, 
 the $J=1$ part of $\Gamma_{pn}$ is absent.
As the $\Gamma_{pn} (J=1)$ is the largest contribution of OPE due mainly 
 to the strong tensor force, the OPE predicts very small total 
 nonmesonic decay rate.  
The kaon exchange and DQ enhances $\Gamma_{nn}$, and the total 
 decay rate reaches up to $0.19$.
Thus the total nonmesonic decay rate of ${^4_\Lambda}$H 
 is a good indicator of the non-pion contributions to the decay process.

\begin{table}[tb]
\caption{Nonmesonic decay rates of ${^4_\Lambda}$H
         in unit of $\Gamma_\Lambda$ }
\begin{center}
\begin{tabular}{p{2cm}p{2cm}p{2cm}p{2cm}p{2cm}p{2cm}}
\hline
${^4_\Lambda}$H
& total & $\Gamma_{pn}$ & $\Gamma_{nn}$ &
$\Gamma_{nn} / \Gamma_{pn}$  & PV/PC \\
\hline
$\pi$      & 0.040 & 0.011 & 0.029 & 2.597 & 7.953 \\
$\pi$+$K$    & 0.071 & 0.005 & 0.067 &14.225 & 6.882 \\
DQ         & 0.040 & 0.013 & 0.027 & 2.028 & 4.238 \\
$\pi$+$K$+DQ & 0.187 & 0.030 & 0.157 & 5.318 & 8.622 \\
\hline
~\ $\beta = +0.1$
           & 0.205 & 0.025 & 0.181 & 7.171 & ----- \\
~\ $\beta = -0.1$
           & 0.168 & 0.034 & 0.134 & 3.938 & ----- \\
\hline
EXP \cite{Out:NPA}
           & 0.17$\pm$0.11 & -----
                           & ----- & ----- & ----- \\
\hline
\end{tabular}
\end{center}
\end{table}

\subsection{Breaking of the $\Delta I = 1/2$ rule}

It is worthy to note that our model does not predict $\Delta I =1/2$ 
dominance.
In the present nonmesonic weak decays,
the $\Delta I =1/2$ rule requires
\begin{equation}
   \left. \matrix{ a_{nn} \cr b_{nn} \cr f_{nn} \cr} \right\}
   = \sqrt{2}
   \left\{ \matrix{ a_{pn} \cr b_{pn} \cr f_{pn} \cr} \right.
\end{equation}
between the amplitudes in Table \ref{TAB:amp}.
It is, however, strongly violated in the $J=0$ amplitudes, i.e., 
$a$ and $b$.
The amplitudes calculated in our $\pi+K+DQ$ model are given in Table 
 \ref{TAB:di12}.
Since we use the $\Delta I =1/2$ rule at the $\Lambda \to N \pi$
 and $\Sigma \to N \pi$ vertices,
 the relations are satisfied in the $\pi + K$ model.

However, in the $\pi + K + DQ$ model,
 one sees that the $\Delta I =1/2$ relations are largely violated
 in $J=0$ amplitudes, while they are almost maintained in the $J=1$
 channels.
Of course this violation comes from the DQ contribution.
Because the $J=0$ amplitudes are relatively small in magnitude,
 their $\Delta I =1/2$ breaking does not show up
 in the decay rate of ${^5_\Lambda}$He.
The ratio of the average $\Gamma_{nn}$ and $\Gamma_{pn}$ is not
 affected much by the $\Delta I = 3/2$ components.
Therefore the large $\Gamma_{nn}/\Gamma_{pn}$ ratio predicted in the DQ
 model is independent of the breaking of the $\Delta I =1/2$
 dominance.

In order to check whether the $\Delta I = 1/2$ dominance is realized
 in the nonmesonic decays,
 the selectivity in the $A=4$ system is again useful.
In fact, we can write
\begin{eqnarray}
\kappa \equiv 
 \frac{\Gamma_{nn}({^4_\Lambda}{\rm{He}})}{\Gamma_{pn}({^4_\Lambda}{\rm{H}})}
 \sim
 \frac{|a_{nn}|^2 + |b_{nn}|^2}{|a_{pn}|^2 + |b_{pn}|^2} .
\end{eqnarray}
 to the leading order.
This ratio should be $2$ if the $\Delta I= 1/2$ transition is dominant.
We can test the $\Delta I = 1/2$ dominance in the $J=0$ amplitudes 
 $a$ and $b$ by seeing the deviation of $\kappa$ from $2$.
Our $\pi + K + DQ$ model predicts $\kappa = 0.133$ reflecting 
 the large $\Delta I = 3/2$ contribution.
Unfortunately, we cannot confirm our result because 
 no experimental data is available for $\Gamma_{pn}({^4_\Lambda}{\rm{H}})$
 so far.
Further experimental studies are highly desirable.
\begin{table}
\caption{The $a$, $b$, and $f$ amplitudes for $A=5$ system in
 arbitrary units.}
\begin{center}
\begin{tabular}{cc|ccc|ccc|}
 & & \multicolumn{3}{c|}{$\pi$+$K$} & \multicolumn{3}{c|}{$\pi$+$K$+DQ} \\
\cline{3-8}
  & & $pn$ & $nn$ & $nn/pn$ & $pn$ & $nn$ & $nn/pn$ \\
\hline
$J=0$ & $a$ & 0.061 & 0.086 & $\sqrt{2}$ & 0.122 & 0.023 & 0.189 \\
      & $b$ & 0.015 & 0.021 & $\sqrt{2}$ & 0.088 & 0.042 & 0.477 \\
\hline
$J=1$ & $f$ & 0.211 & 0.298 & $\sqrt{2}$ & 0.332 & 0.465 & 1.401 \\
\hline
\end{tabular}
\end{center}
\label{TAB:di12}
\end{table}

\section{Conclusions}
We calculate the nonmesonic decay rates of $\Lambda$ hypernuclei
 by using the quark-meson hybrid model.
We have found that the $J=1$ part of $\Gamma_{pn}$ 
 is reduced by the $K$ exchange
 contribution, and the $J=1$ part of $\Gamma_{nn}$ 
 is enhanced both by the $K$ exchange and the DQ contribution.
Thus the $\Gamma_{nn}/\Gamma_{pn}$ ratio becomes large,
 and the result is consistent with the current experimental data.

We also estimate the effect of the virtual $\Sigma$ excitation
 which is known to be important when we consider the property of
 $\Lambda$ in nuclei.
In this paper 
 we add the $\Sigma N \to NN$ amplitude to the $\Lambda N \to NN$
 amplitude by using the mixing parameter $\beta$.
We find that it changes the magnitudes of the nonmesonic decay 
 rate significantly and our estimates of the decay rates agree with 
 the experimental data fairly well for a negative $\beta \simeq -0.1$.
We have also pointed out that 
 observation of final $p$-$p$ state from the decay of ${^4_\Lambda}$He
 gives us a chance to show a clear evidence 
 of the $\Sigma$ component in $\Lambda$ hypernucleus.

Our model predicts that the ``$\Delta I = 1/2$ rule'' is largely
 violated in the $J=0$ transitions.
In order to confirm the prediction, a careful experiment of the 
 ${^4_\Lambda}$H decay is indispensable, which is now underway.
The origin of the $\Delta I = 1/2$ dominance in nonleptonic 
 $ |\Delta S| = 1$ weak transitions is still under heavy discussion.
The vertex corrections plus the Penguin diagrams were shown to
 enhance $\Delta I = 1/2$.
They alone, however, are not enough to reproduce the large ratio
 of the $\Delta I = 1/2$ and $\Delta I = 3/2$ amplitudes.
There must be significant effects from the nonperturbative QCD
 corrections.
The $\Lambda N \to NN$ transition will be a new tool to determine the 
 nonperturbative mechanism of the $\Delta I = 1/2$ enhancement.

\section*{Acknowledgments}
The authors thank Prof. A. Gal, Prof. Y. Akaishi, Dr. E. Hiyama, 
 and Prof. M. Iwasaki for discussions.
This work is supported in part by the Grant-in-Aid for Scientific Research
 (C) (2) 11640261 of Japan Society for the Promotion of Science.

\end{document}